% MNSAMPLE.TEX
%
% A sample plain TeX single/two column Monthly Notices article.
%
% v1.5  --- released 25th August 1994 (M. Reed)
% v1.4  --- released 22nd February 1994
% v1.3  --- released  8th December 1992
%
% Copyright Cambridge University Press

% The following line automatically loads the mn macros if you are not
% using a format file.
\ifx\mnmacrosloaded\undefined \input mn\fi

% If your system has the AMS fonts version 2.0 installed, MN.tex can be
% made to use them by uncommenting the line: %\AMStwofontstrue
%
% By doing this, you will be able to obtain upright Greek characters.
% e.g. \umu, \upi etc.  See the section on "Upright Greek characters" in
% this guide for further information.

\newif\ifAMStwofonts
% \AMStwofontstrue

\ifCUPmtplainloaded \else
  \NewTextAlphabet{textbfit} {cmbxti10} {}
  \NewTextAlphabet{textbfss} {cmssbx10} {}
  \NewMathAlphabet{mathbfit} {cmbxti10} {} % for math mode
  \NewMathAlphabet{mathbfss} {cmssbx10} {} %  "   "    "
  \ifAMStwofonts
    \NewSymbolFont{upmath} {eurm10}
    \NewSymbolFont{AMSa} {msam10}
    \NewMathSymbol{\upi}     {0}{upmath}{19}
    \NewMathSymbol{\umu}     {0}{upmath}{16}
    \NewMathSymbol{\upartial}{0}{upmath}{40}
    \NewMathSymbol{\leqslant}{3}{AMSa}{36}
    \NewMathSymbol{\geqslant}{3}{AMSa}{3E}

    \let\leq=\leqslant 
     \let\ge=\geqslant
  \else
    \def\umu{\mu}
    \def\upi{\pi}
    \def\upartial{\partial}
  \fi
\fi

% Marginal adjustments using \pageoffset maybe required when printing
% proofs on a Laserprinter (this is usually not needed).
% Syntax: \pageoffset{ +/- hor. offset}{ +/- vert. offset}
% e.g.    \pageoffset{-3pc}{-4pc}

\pageoffset{-2.5pc}{0pc}

\loadboldmathnames

% \Referee   %  uncomment this for referee mode (double spaced)

% \pagerange, \pubyear and \volume are defined at the Journals office and
% not by an author.

% \onecolumn        % enable one column mode
% \letters          % for `letters' articles
% \pagerange{1--7}    % `letters' articles should use \pagerange{Ln--Ln}
% \pubyear{1989}
% \volume{226}
% \microfiche{}     % for articles with microfiche
% \authorcomment{}  % author comment for footline

\begintopmatter  % start the two spanning material

\title{Dynamical friction in dwarf galaxies}
\author{X. Hernandez$^{1,2}$ and Gerard Gilmore$^1$}
\affiliation{$^1$ Institute of Astronomy, Cambridge University, Madingley Road, Cambridge CB3 0HA}
% \affiliation{$^2$ Instituto de Astronom\'\i a, Universidad Nacional Aut\'onoma de M\'exico, A.P. 70-264, 04510 M\'exico, D.F.}

\shortauthor{X. Hernandez and G. Gilmore}
\shorttitle{Dynamical friction in dwarf Galaxies}

% \acceptedline is to be defined at the Journals office and not
% by an author.

% \acceptedline{Accepted 1988 December 15. Received 1988 December 14;
%   in original form 1988 October 11}

\abstract {
We present a simplified analytic approach to the problem of
 the spiraling of a massive body 
orbiting within the dark halo of a dwarf galaxy. This dark halo is
treated as the core region of a King distribution of dark matter
particles, in consistency with the observational result of dwarf
galaxies having solid body rotation curves. Thus we derive a simple
formula which provides a reliable and general first order solution to
the problem, totally analogous to the one corresponding to the
dynamical friction problem in an isothermal halo. This analytic
approach allows a clear handling and a transparent understanding of
the physics and the scaling of the problem. A comparison with the
isothermal case shows that in the core regions of a King sphere,
dynamical friction proceeds at a different rate, and is sensitive to
the total core radius. Thus, in principle, observable consequences may result.
In order to illustrate the possible effects, we apply this formula to the spiraling of
globular cluster orbits in dwarf galaxies, and show how present day
globular cluster systems could in principle be used to derive better limits on the
structure of dark halos around dwarf galaxies, when the observational
situation improves. As a second application, we study the way a
massive black hole population forming a fraction of these dark halos would
gradually concentrate towards the centre, with the consequent
deformation of an originally solid body rotation curve. This effect
allows us to set limits on the fraction/mass of any massive black hole minority
component of the dark halos of dwarf galaxies. In essence, we take
advantage of the way the global matter distribution fixes the local
distribution function for the dark matter particles, which in turn
determines the dynamical friction problem.
}


\maketitle  %  finish the two spanning material

\section{Introduction}

Over the past few years it has become apparent that dwarf galactic
systems distinguish themselves from the larger class of galaxies in
more ways than their total luminosity. Rotation velocity studies in
dwarf spirals and velocity dispersion studies in dwarf irregulars and
dwarf spheroidals have revealed the presence of a dark, dynamically
dominant component, much in excess of what a scaled down version of a
large galaxy would show. The dark matter halo of a dwarf galaxy is
more massive than an extrapolation of a single mass/luminosity
relationship fitted to massive galaxies would suggest (eg, that of
Kormendy 1986), but it also appears to have a distinct type of
structure.

One of the best-studied dark matter distributions is that of the
Sagittarius dSph galaxy (Ibata etal 1997). Ibata etal (1997) analyzed
their kinematics of this galaxy to derive a (radial) period of the
orbit of Sgr about the Galaxy of less than $\sim 1$Gyr.  This limit,
together with lifetime constraints derived from the ages of the old
stellar populations in Sgr, show it has survived for more than 10
orbits. Numerical models of the survival of a dSph galaxy orbiting
inside a larger Galactic halo, however, imply complete tidal
disruption in a few orbits (Velazquez and White 1995; Johnston etal
1995). The solution of this paradox suggested by Ibata etal is that
the dark matter halo of Sgr, whose existence is derived directly from
their kinematics, must have an approximately constant (Heaviside )
density profile, out to a cutoff at the tidal radius. This model is
consistent with all available constraints. The rotation curves of
gas-rich dwarf spirals also typically show `solid-body' linear behavior
in their inner few kpc (eg, Casertano \& van Albada 1990 or Burkert 1995).
Other examples of galactic systems where observations suggest the presence of
a constant density structure in the central regions are the
rotation curves of de Blok et al. (1996) for LSB's and  
Pryor \& Kormendy (1990) for dSph density profiles.
Additionally, there is the case of DDO 154, a nearby dwarf galaxy 
the rotation curve of which has been extensively studied in detail. 
This galaxy shows a core region, a maximum, and then an almost Keplerian
decline, in substantial disagreement with models of dark halos obtained in
cosmological simulations, as
analyzed by Burkert \& Silk (1997).
These observational results are in disagreement with the singular
profiles obtained in self-consistent cosmological simulations (e.g. Navarro et al. 1996)
so that the detailed structure of dark matter halos is presently the object of some
debate e.g. Burkert (1997). 
From a theoretical point of view, in our paper I (Hernandez \& Gilmore 1997)
we use a simple baryonic infall model to infer the initial dark halo
profiles of late type galaxies. Using the recently observed rotation curves
of these systems (de Blok et al. 1996) and other observational restrictions
to calibrate the initial halo profiles, we show that a King profile can accurately
reproduce the observed rotation curves of these systems, as well as those of normal
late type galaxies. In that paper we also show that centrally divergent density profiles
are difficult to reconcile with the rotation curves of LSB galaxies. 
In as much as the problem is not settled, we explore the case of
a constant density halo for dwarf galaxies, (as observations suggest) in terms of
its possible implications for dynamical friction in those systems.

That is, in general, whereas more luminous spiral galaxies show a
typically flat rotation curve out to several disk scale radii,
indicating isothermal halos with density falling radially as $r^{-2}$,
dwarf galaxies appear to have approximately constant density halos out
to the edge of the stellar distribution. Moreover, while the central
regions of luminous galaxies are gravitationally dominated by stars,
so that inferences about the inner structure of the dark matter are
necessarily model-dependent, the central regions of dwarf galaxies are
gravitationally dominated by dark matter. In this paper we consider
the consequences of this relatively flat mass density profile for
dynamical evolution in the dwarf galaxy, to consider the possibility 
that this dominance by dark matter can be exploited to
constrain the nature of that dark matter, particularly by exploiting
dynamical friction.

Dynamical friction has important consequences in shaping the orbital
evolution of massive bodies within dark halos, such as globular
clusters and any hypothetical massive black holes. This has been used
to set some interesting constraints on the parameters of the massive
objects which orbit in the dark halo systems of large galaxies, such
as the allowed masses of black holes, required to be consistent with
dynamical friction not having made their orbits decay over their
lifetimes e.g. Hut \& Rees (1992).  Similarly, if the orbiting objects
are better known, as in the case of globular clusters, the problem can
be inverted, and used to set limits on the dark halo structure and the
evolution of the globular cluster system (eg Aguilar, Hut and Ostriker
1988). Nothing which is not derivable from the complete form of the
rotation curve can be derived about the distribution function of the
dark matter itself. In the case of dwarf systems however, which remain
dark matter dominated into their inner regions, we can use dynamical
friction constraints to investigate the distribution function of the
dark matter in the core region, determined by the total matter
distribution, even though only a fraction of the rotation curve may be
accessible to observations. It is therefore interesting to use
dynamical friction constraints in the case of the dark matter
dominated dwarf galaxies, to try to learn more about the dark halos in
these systems.

Although the problem reflects the gravitational interactions of the
massive body with the totality of halo particles, and strictly should
be treated through n-body simulations, a few well grounded assumptions
can simplify the problem to allow an analytical treatment. This has
been done (e.g. Binney \& Tremaine 1987) in the case of the isothermal
halo, to provide a robust, general purpose analytical formula which
can give a reliable first order solution, which although not as
accurate as an n-body code, has the advantage of providing physical
insight into the solution and the scaling of the problem. Dynamical
friction is most sensitive to the low velocity particles, which are
the ones which interact more strongly with the body undergoing this
friction. The presence of a core in a dark halo means effectively
changing the ratio between low and high velocity halo particles,
towards more low velocity ones, as the high energy particles are
mostly found on extended orbits. Hence, it seems reasonable to suspect
that a distribution function corresponding to a system with a core
will change the problem of orbital spiraling due to dynamical friction
with respect to the case of an isothermal dark matter halo. Since dSph
galaxies do have globular clusters in some cases (Sgr has four) one is
interested to know if their continued existence provides any useful
limits on the dark matter. In this
paper we develop a formula totally analogous to the one of Binney \&
Tremaine (1987) for the gravitational friction on a massive body
orbiting within an isothermal sphere, but appropriate to the dwarf
galaxy dark matter problem.

In this paper the dynamical friction formulation for the core regions
of dwarf galaxies is derived and applied in connection with two
aspects of the dynamics of these systems. First, we try to provide
information on the structure of these dark halos, from regions beyond
those accessible through direct measurements of stellar velocities,
using the global information contained in the local distribution
function of the halo particles, through dynamical friction
effects. The orbital spiraling of a globular cluster will depend on
the global dark halo structure, as it is this which determines the
local distribution function, which in turn determines the dynamical
friction problem. We can find what range of halo structures is
consistent with some observed globular cluster system age and orbital
distribution. Second, we consider a more general case, where some part
less than 100 percent of the dark halo is in the form of massive
compact objects (black holes?), and derive some limits on the mass of
any such black holes by requiring that their orbits should not have
decayed over the lifetime of the system.

In dwarf galaxies, it can not be expected that the density profile
inferred from the regions where the stars can be measured extends
indefinitely in a shallow density profile.  It seems more natural that
we should be seeing only the core region of a distribution of dark
matter, the rest of the halo being empty of stars, see for example
Pryor \& Kormendy (1990) and Lake (1990). King-model spheres represent
a self-consistent solution to both a Boltzmann and a Poisson equation,
and have been found to fit well the end products of N-body violent
relaxation simulations, so we adopt here the plausible simplifying
assumption that we may treat galactic dark halos as King spheres.
We note that in our paper I we find King profiles to adequately
reproduce the rotation curves of LSB and dwarf galaxies.
In that paper we perform a detailed study of baryon dissipation 
within dark halos, calibrated using a variety of observational relations.

In Section 2) we present the derivation of the simple analytical
formula describing orbital spiraling due to dynamical friction in
the core regions of dark halos, which we apply in section 3) to the
two problems mentioned above, in connection to dwarf galaxies. 
In section 4) we present the conclusions of this paper.

\section{Theoretical approach}

Take a body A of mass $M$ moving on a stable circular orbit around the
centre of a spherically symmetrical mass distribution $\rho(r)$, made
up of an equilibrium distribution of self gravitating particles of
mass $m$, where $M>m$.  Consider one gravitational encounter between the
body A and one of the background particles. This produces a small
deviation in A's orbit producing a negative change $-\Delta V$ in the original
forward velocity, V, and a positive change $+\Delta V$ perpendicular to the
original forward velocity, in the plane of the interaction. Consider
now the totality of the background particles. In this case all the
$+\Delta V$ deviations will average out to zero, and only a net
$\Sigma (-\Delta V)$ will remain, in the direction of the original
velocity.  This deceleration is what constitutes dynamical friction. 
 
If one now integrates over the effects of particles interacting with impact
parameters from 0 to $b_{max}$ , a maximum impact parameter relevant
to the problem, and assumes that the background particles move
isotropically, one obtains: 
$$
   {dV \over dt} =-16 \pi^2 {ln\Lambda G^2 m M \over V^2} \int_{0}^{V}
   f(v)v^2 dv, \eqno\stepeq 
$$
for the deceleration parallel to V experienced by A as a result of the
collective interactions with the background particles having a
distribution function $f(v)$, isotropic in velocity space. In equation
(1) $\Lambda = {b_{max}V_T^2 /(GM)}$, and $V_T$ is a typical velocity
of the system (see Binney \& Tremaine 1987). It has been assumed that
$ln \Lambda$ is a constant, although this is not strictly true, but
usually $\Lambda >> 1$, and $ln \Lambda$ does not vary appreciably for
most applications. Similarly, $ln \Lambda $ is not sensitive to the
choice of $V_T$, taken in this work as the circular velocity of the
halo of background particles. Numerical experiments have confirmed the
validity of (1), and of $ln \Lambda=cte.$ e.g. Bontekoe \& van Albada
(1987) and Zaritsky \& White (1988), see Binney \& Tremaine (1987) for
a more detailed discussion, and a derivation of equation (1).
  
As explained in the introduction, in order to use equation (1) in the
case of dwarf galaxies, we shall assume their dark matter halos to be
well represented by the core regions of a King distribution,
approximated by a constant density region, i.e.,
$$
   \rho (r<r_0) =\rho_0   \eqno\stepeq
$$
and 
$$
   f(v)= {n_0 \over (2 \pi \sigma^2)^{3/2}} \left( exp(-v^2/ {2
   \sigma^2})-exp(-v^2_e/ {2 \sigma^2}) \right), \eqno\stepeq
$$

In equation (2) $\rho_0$ is the density within the core region, $r_0$ the
core radius, defined as $r_0 = \left( 9 \sigma^2/ 4 \pi G \rho_0
\right)^{1 \over 2}$, and $\sigma $ is the isotropic velocity
dispersion of the halo particles. In equation (3), $n_0$ is a particle
number density, $n_0 ={\rho_0/m}$, and $v_e =v_e (r) $ is the escape
velocity of the halo. Introducing equation (3) into equation (1), one
obtains:
$$
   {dV \over dt} =-16 \pi^2 {ln\Lambda G^2 \rho_0 M I(r)V^{-2}},
   \eqno\stepeq 
$$ 
where, 

$$
I(r)={1 \over (2 \pi \sigma^2)^{3/2}} \int_{0}^{V} \left( exp({-v^2
\over {2 \sigma^2}})-exp({-v^2_e \over {2 \sigma^2}}) \right) v^2 dv. 
$$

Using the substitutions $X={v/(\sqrt{2} \sigma)}$, $Y={v_e/
(\sqrt{2} \sigma)}$ and $X_A ={V \over (\sqrt{2} \sigma)}$, $I(r)$
becomes: 

$$
I(r)={1 \over \pi^{3/2}} \int_{0}^{X_A} X^2 \left( e^{-X^2}-e^{-Y^2}
\right) dX. 
$$

$$
   \Longrightarrow I(r)={1 \over 4 \pi} \left( erf(X_A)-{2 X_A \over
   \sqrt{\pi}} e^{-X_A^2}-{4 X_A^3 \over 3 \sqrt{\pi}} e^{-Y^2}
   \right) \eqno\stepeq 
$$
 
At this point we have to introduce an assumption about the orbit of
body A, so that we can evaluate $X_A$.  Since the dynamical friction
drag removes energy from A in proportion to $(dV)^2$, and angular
momentum only in proportion to $dV$, as time progresses, A will tend
to settle into an orbit of maximum angular momentum. This process
leads to the circularisation of the orbits of bodies under the
influence of dynamical friction. In this work, we will assume that the
orbits of spiraling bodies are always circular, (see for example Wahde
\& Donner (1996), who study the more difficult problem of the
influence of the disk on the dynamical friction problem, also under
the assumption of circular orbits for the spiraling body) in
accordance with the objective of deriving a simple formula for the
process. In this case, $V=V_c$, where:

$$
V_c^2 (r)={GM_H(r) \over r}= {4 \pi \over 3} G \rho_0 r^2 
$$

$$
\Longrightarrow V_c(r)=V_0(r/r_0) 
$$
Now $V_c(r_0)=V_0$, and $X_c={V_c /\sqrt2 \sigma} \equiv X$, and
$M_H(r)$ refers to the dark halo mass internal to radius $r$, with
$M_H(r_0)=M_H$, the total halo mass within the core radius. 

Now we need $v_e$, the escape velocity. From equation (2), 

$$
{1 \over 2}v_e^2(r)= \int_{r}^{r_0} {G M_H(r) \over r^2} dr +
\int_{r_0}^{\infty} {G M_H \over r^2} dr . 
$$
The second integral is an underestimate, as $\rho(r>r_0) \ne 0$
(except in a tidally truncated dSph, such as Sgr), but since we are
interested only in the region $r<r_0$, which corresponds to $X<1.23$,
this underestimate introduces little error-- notice the $X^3$ and the
$exp({-Y^2})$ in the relevant term in I(r). Now,

$$
{1 \over 2} v_e^2(r)={4 \pi \over 3} G \rho_0 \left( {{r_0^2-r^2 \over
2} + r_0^2} \right) 
$$

$$
   v_e^2(r)=V_c^2(r) \left( {3r_0^2 \over r^2}-1 \right) \eqno\stepeq
$$

$$
\Longrightarrow Y^2=X^2 \left( {3r_0^2 \over r^2}-1 \right) 
$$
Introducing $R=r_0/r$, $V_c(R)=(3^{1/2} \sigma R)$ and
$Y^2=(9/2-X^2)$ : 

$$
   I(X)=
$$
$$
\hskip 8mm {1 \over 4 \pi} \left( erf(X)-{2X \over \sqrt{\pi}}exp({-X^2}) -{4X^3 \over 3 
	\sqrt{\pi}}exp({X^2-9/2)} \right). \eqno\stepeq 
$$

Now to calculate the spiraling of A, we use the assumption that it
always orbits at $V_c$, therefore, $\left| L \right| =r M V_c$ and 
$$
{d(L/M) \over dt}=V_c {dr \over dt} 
$$
Also, the acceleration in equation (4) corresponds to a drag force $M(dV_c
/dt)$ parallel to $V_c$, and therefore a torque 
$$
{d(L/M) \over dt}=r{dV_c \over dt} 
$$

$$
   \Longrightarrow {dr \over dt}= {r \over V_c}{dV_c \over dt} \eqno\stepeq
$$

Now define $\tau$, the characteristic time-scale with which the orbit
of A decays as: 
$$
\tau = {r \over 2(dr/dt)_r} 
$$
using  equation (8), $$\tau= {V_c \over 2(dV_c /dt)_r}$$ and from equation (4), 
$$
\tau ={ V_c^3 \over 32 \pi^2 ln\Lambda G^2 M \rho_0 I(X)} 
$$
Substituting $V_c$ for $\rho_0$, $V_0$ for $V_c$ and X for r, we get: 

$$
   \tau =\left( V_0 \over r_0 \right) \left( 2 \over 3 \right)^{3/2}
   {r_0^3 \over 24 \pi ln\Lambda M G} \left( X^3 \over I(X) \right) \eqno\stepeq
$$

Now if $F(X)= X^3/I(X)$, we see in Fig. 1 that $F(0)=16.89$, remaining almost
constant as $X$ increases, reaching 19 for $X=0.5$, and increasing
slightly to reach 29 at $X=1$, and increasing slightly faster
beyond that point. This
shows that orbital decay in the core region of a King sphere from
$R=1$ to around $R=0.8$ proceeds faster than exponential, and after
that it becomes essentially exponential, with time-scales: 
$$
\tau =\left( 20 \right) \left( V_0 \over r_0 \right) \left( 2 \over 3 \right)^{3/2}
{r_0^3 \over 24 \pi ln\Lambda M G},  
$$
which comes to: 

$$
   \tau_{DF}=\left( V_0 \over r_0 \right) {r_0^3 \over 3 M ln \Lambda}
   Gyr, \eqno\stepeq
$$
which is the final result of this section, where $[V_0]=km/s$,
$[r_0]=kpc$ and $[M]=10^5 M_{\odot}$. The main differences between this
result and that for the isothermal case in Binney \& Tremaine is
the density profile, and the fact that $v/ \sigma$ is a function of
$r$, rather than $X \equiv 1$. This yields an exponential time-scale
independent of the radius at which the particle orbits and which is only
a function of the central density of the halo, and the total core
radius.

\beginfigure{1}
\vskip 91mm
\caption{{\bf Figure 1.} The function involved in equation (9), 
$F(X)$ and the more rigorously derived function $F'(X)$ described in
Appendix B, for $0<X<1$, the range corresponding to $R \leq 0.82$.
Notice the lack of any strong variations in $F(X)$ over this range,
which has been used to approximate this function by a constant, close
to the value it takes in the inner regions. The more rigorously
correct expression $F'(X)$ is seen not to differ much from $F(X)$,
showing that the escape velocity approximation introduces little
error.}
\endfigure

\subsection{Comparison with the isothermal case}

At this point, it is interesting to compare equation (10) with the
corresponding expression describing the spiraling of a body of mass M
moving on a circular orbit around the centre of an isothermal density
distribution, characterized by a constant circular velocity $V_c$, 
$$
   t_{DF} = {2.64 r_i^2 V_c \over M ln \Lambda} Gyr, \eqno\stepeq 
$$
(Binney \& Tremaine 1987), where $t_{DF}$ is the total time it takes
for the body to spiral from an initial radius $r_i$ to $r=0$, and
$\Lambda$ and the units are the same as in equation (10). The main
difference between the two cases, is that in the isothermal one the
spiraling body reaches the centre in a finite amount of time, whereas
in the core region case the orbital spiraling only asymptotically
reaches the centre, with a half-orbit time, $\tau_{DF}$.

To compare these two results, consider an observed body at a
galactocentric distance of $r_B$, orbiting at a velocity $v_B$. We
wish to compare the evolution of its orbit in an isothermal sphere,
which is an adequate first approximation to the outer parts of a dark
halo in at least a large galaxy, to the evolution of its orbit in a
constant density core, which is the topical model for the inner
regions of dwarf galaxies. For the constant density core,
$$
   V_0=v_B \left( r_0 \over r_B \right)
$$
and therefore, from equations (10) and (11),
$$
{\tau_{DF} \over t_{DF}} =\left( 1 \over 8 \right) \left( {r_0 \over
r_B } \right) ^3
$$
In practice, the inner regions of large galaxies are baryon-dominated
rather than dark-matter dominated, and even in dwarfs the baryon
component is not always negligible, so that interactions with the
baryonic component, which increase as the radius decreases, will
substantially modify the orbit at late times. To avoid this
unnecessary complication, we take four
half-orbit times as representative of the orbital decay in the core
region case. This fiducial number of half lives is only used for this
particular comparison with the isothermal case, and is not used again
in any of what follows. After this time the body will have decayed into an orbit
with a radius $1/16$ of the initial radius. In this case,
$$
{{4 \tau_{DF}} \over t_{DF}} =\left( 1 \over 2 \right) \left( {r_0 \over
r_B } \right) ^3
$$
We see that for the simplest assumption of $r_B = r_0$, dynamical
friction decay timescales are twice as fast in the case of a constant
density core than in the case of an isothermal distribution. That is,
the effects can be relatively large, enough to be observable. 
For $(r_0 / r_B) = 2^{1/3}$ the total orbital
decay timescale is equal in both cases, and becomes progressively
longer in the constant density case, as the true core radius becomes
larger than the observed orbital radius. Whereas in the isothermal case the observed
radial distance and orbital velocity of the body uniquely determine how long the dynamical
friction process will take to drive the body to $ r=0 $, in the core region case one
requires the central density ( a velocity radius pair within the region of interest) and
the halo core radius. In the isothermal case equation (11) yields a fixed $t_{DF}$
for a body with an observed radial distance and rotation velocity, while equation (10)
further requires an assumed core radius for the system, which could sometimes be inferred
from consistency requirements (see the case of the Sgr dwarf below).

It is this dependence of the orbital decay process on the total core radius
which could be used to derive structural halo parameters from orbital
structures. In the isothermal case, only the starting radius
determines the spiraling process: there is no sensitivity to the
global parameters. Presented differently, the only parameter of the isothermal
halo is $V_c$. It should be noted that the comparison between the two
time scales is highly sensitive to $r_0 /r_B$ , and therefore, the
isothermal formula is in general not a reliable estimate of the
dynamical friction problem in cases where a constant density core is
suspected. Additionally, in some applications the actual evolution of
the orbit might be relevant, in which case the better fitting of
equation (10) or equation (11) might reveal the presence of a core
region in the dark halo in question.
 
Finally, it should be noted that neither of equation (10) nor (11) can
be applied to the case of the orbital decay of dwarf spheroidal
galaxies in the halo of our galaxy.  Direct application of equation
(10) would assume that only the halo matter is present, so that the
baryonic component of our galaxy, which has an important dynamical
contribution over any possibly interesting dark halo core region,
would be ignored. This would exclude an important fraction of the
rotation velocity from consideration, resulting in erroneously short
times being predicted. Equation (11) assumes that all the matter
responsible for the flat rotation curve contributes to the dynamical
friction drag, which would again be an overestimate of the dynamical
friction, as the disk material does not interact with the dwarf galaxy
in the same way as the halo particles.  Additionally, one would
require a consistent distribution function for the halo particles, in
the presence of dynamically important disk and bulge components. For
the above reasons, equation (10) is only relevant to the internal
dynamics of dwarf galaxies, or other clearly dark matter dominated
systems, where a uniform radial density is suspected from
observations. Such cases do exist (Ibata etal 1997), and may even be the
norm with LSB galaxies (paper I).

\section{Applications}

In this section we use equation (10) in two interesting problems, to
present a theoretical framework which should allow us to derive
important constraints on the structure of the dark halos of dwarf
galaxies, and to set some constraints on the fraction of these halos
which could be made up of massive compact objects.

\subsection{Core radii determination in dwarf galaxies}

We can use equation (10) to obtain information on the size of the core
region of dwarf galaxies by considering the effects of dynamical
friction on the globular cluster systems of these galaxies.  It is of
course notoriously difficult to deduce the properties of a
hypothetical destroyed parent population from few, or no,
survivors. Nonetheless, our aim here is to illustrate the sensitivity
of such analyses to assumptions made concerning the spatial density
distribution of the dark matter, independent of current observational
limitations. 

For simplicity here, to illustrate the scale of the effect
to zeroth order, we characterize luminous globular clusters as
having uniform parameters, in particular masses close to $M=10^5
M_{\odot}$ ( see Harris 1991 for a review on the subject). Taking
$M=10^5 M_{\odot}$, representative of a typical globular cluster,
$b_{max}=3 kpc$, in the range of the solid body rotation regions of
dwarf galaxies, and $V_T=40 km/s$, used only to calculate $ln
\Lambda$, which is only marginally sensitive to these values, we
obtain:
$$
   \tau_{DF}= \left( {V_0 \over r_0} \right) {r_0^3 \over 27.8 M}
   Gyr.  \eqno\stepeq 
$$

Suppose that a dwarf galaxy is observed, having a solid body rotation
curve out to the last measured point, at $r=2.5 kpc$, with $V_c(2.5
kpc)=30 km/s$ (typical values for these systems, for example Carignan
\& Beaulieu (1989) for the case of DDO 154 in which case the HI
rotation curve was measured beyond the extent of the stellar content).
The null hypothesis would be to assume that the core region of this
galaxy measures $2.5 kpc$, extending only as far as the rotation curve
could be measured, with $V_0=30 km/s$, but this is clearly only a
lower limit.  If we take $r_0 =2.5 kpc$ and $V_0 =30 km/s$, for
globular clusters of $M=10^5 M_{\odot}$, equation (12) gives:
$$
\tau_{DF} = (30 / 2.5)(2.5^3 / 27.8) = 6.74 Gyr.
$$ 

This timescale is sufficiently short compared to the ages of dwarf galaxies
in the Local Group as to be potentially interesting. In general,
as the dwarf galaxies correspond to higher contrast initial
fluctuations, they became bound structures, and perhaps initiated 
star formation, earlier than normal, larger
galaxies. Additionally, direct stellar population studies in these
systems have yielded population ages of about the age of the universe
(e.g. Hodge 1989, Ibata etal 1997).  In view of the above, $12 Gyr$ seems like a
suitable age for these systems. The application of equation (12) shows
that if the core radius measured only $2.5 kpc$, almost 2 orbit
half-times have elapsed for the globular cluster system, which should
therefore show significant dynamical friction effects. Specifically,
one expects a low abundance of globular clusters at large radial distances
and to find them only concentrated very
close to the centre of the system, where dynamical friction would drive them. 

Were such a distribution observed, could one in fact reliably infer
that dynamical friction might have been to blame?  Suppose instead,
that the actual dark halo core region is $3.5 kpc$ in size. In this
case, as the density of the dark matter would not change, $(V_0 /r_0)$
remains the same, and we obtain,
$$
\tau_{DF}= 6.74{(3.5/2.5)}^3 = 18 Gyr.
$$
This last value is larger than the age of the universe, and we would
therefore expect to see no dynamical friction effects in the globular
cluster system of this galaxy. 

Clearly, the present spatial distribution of globular clusters
in dwarf galaxies is a (one-way) test of the importance of dynamical
friction, by investigating the incidence of spatially extended cluster
systems in galaxies with dark-matter dominated, linear, rotation curves.
This program is not applicable at the present time, as we lack
information on the state and nature of globular cluster systems around
dwarf galaxies. In fact, due to observational difficulties, together
with the intrinsic rarity of globular cluster systems in dwarf
galaxies, there are presently only a handful of detections of globular
clusters around dwarf galaxies. It is through the use of accumulated deep
HST and wide angle photometry and redshift surveys (e.g. SDSS, 2dF)
that this problem might be explored.

As a specific example, we can take the case of the Sagittarius dwarf,
the recently discovered dSph galaxy, whose dark matter distribution,
as derived from stellar kinematics, is in good agreement with the core
model discussed in this paper.  The Sgr dwarf has four globular
clusters, and has been studied fairly well kinematically (Ibata etal
1997).  From the observed velocity dispersion of stars in this
galaxy, we adopt $V(1 kpc)=20 km/s$. Additionally, the tidal radius
for this galaxy at its current position is $\leq 1 kpc$ (Ibata
et. al. 1997).  Applying equation (12) with these numbers, we obtain
$\tau_{DF}=0.72 Gyr$, for globular clusters of $10^5 M_{\odot}$. Since
the globular cluster system of this galaxy has not decayed completely,
as such a short half life would suggest, the Sgr dwarf necessarily had
originally a dark halo core radius of more than its present tidal
radius. The dependence of $\tau_{DF}$ on the total core radius of the
galaxy allows to reconcile theory with observations, as a larger core radius
(not constrained by any direct observation) would result in more extended
values for $\tau_{DF}$.

Quantifying the original size is necessarily inexact. If, for example,
we assume that there have elapsed only $2$ half lives for this
globular cluster system, equation (12) yields an original core radius
of $2.4 kpc$, which is larger than the observed $1 kpc$ tidal radius. 
This is consistent with the finding that the stellar
population which has been associated with this galaxy presently
spreads over $3 kpc$, showing signs of tidal disruption by the
Galaxy. This is what would be expected if the original extent of this
galaxy had been larger than its present tidal radius, as equation (12)
suggests. Further, it being the only dSph with a globular cluster
system also suggest the Sagittarius dwarf had a large original size.
Thus, the existence of a globular cluster system, together with the
constant density dark matter mass distribution derived from stellar kinematics,
and the dynamical friction analysis of this paper,
requires that Sgr is significantly tidally stripped.

Thus, through requiring that the dynamical friction decay time, 
$\tau_{DF}$ be consistent with the spatial extent
of the globular cluster system over the age of the galaxy we have used equation (12)
to set some constraints on the size of the original core region of this system.
If we took the formula for the isothermal halo, we would obtain $t_{DF}=8\tau_{DF}=$5.7 Gyr,
also significantly shorter than the age of this system. Since, in the isothermal case, there is no further
dependence beyond the observed position and orbital velocity, there would be no way to
reconcile theory with observations in this case. This example serves to illustrate the differences
in the dynamical friction problem between the isothermal and constant density core cases, 
as well as to provide independent confirmation for the dynamical calculations of 
Ibata et al (1997), in the sense that the dark halo of the Sgr dwarf is probably
characterized by a constant density profile.

\subsection{The case of massive black holes in the dark halos of dwarf
galaxies} 

In this sub-section we shall use equation (10) to set some limits on
the fractional part of dwarf galactic halos which could be made up of
massive black holes. In the review on the subject of baryonic dark
matter by Carr (1994), it is shown that the only non-excluded baryonic
dark matter candidates for the halos of galaxies (like our own) are
brown dwarf stars and massive black holes in the range $10^3 -10^7
M_{\odot}$. Gravitational microlensing searches (eg Alcock et
al. 1996) are ideal probes at low masses, but are less sensitive to
the very rare and very long time-scale events caused by massive black
holes.  We can provide a limit in the case that massive black holes
make up only a part of the dark halo, by investigating the survival
time of such a system against dynamical friction decay.

We calculate the time it would take for such massive black holes
(making up a fraction $\gamma$ of the total dark halo) to spiral
towards the centre of the dark halos of dwarf galaxies, as a result of
dynamical friction with another dark matter component, made up of
small particles.  To first order, we can expect that after 1 orbit
half-life all the black holes formerly part of the uniform density
distribution within $r_0$ will have formed a new density distribution
within $r_0 /2$, leaving the remaining fraction of dark matter as it
was originally. Clearly, this scenario for the evolution of the halo
is only applicable to low black-hole mass fractions. If the black
holes make up a large fraction of the halo, dynamical friction will
only redistribute energy between the two components, making the
particle distribution expand, with the black holes segregating only
marginally towards the central regions, at which point dynamical
friction would stop operating. In this way, this method is
complementary to the micro-lensing approach, which is not sensitive to
the possibility of only a small fraction of the halo being made up of
massive black holes.

It is easy to show that the fractional increase in the rotation
velocity at a radius $r_0 / 2^n$ after $n$ half-orbit times have
elapsed, and a fraction $\gamma$  (for $\gamma <<1$) of the original constant density halo
has concentrated interior to $r_0 / 2^n$ is:

$$
\mathop{V_c}_{t=n \tau} (r_0 /2^n) = \left( (1- \gamma) +\gamma 2^{3n}
\right)^{1/2} \mathop{V_c}_{t=0}(r_0 /2^n). 
$$

This increment would only appear inwards of $r_0 / 2^n$, the rotation
curve beyond being reduced.  It is this distortion of a solid body
rotation curve, to one with a central enhancement, which would appear
anomalous relative to observed rotation curves, and so provides the
observational application here. In the limiting case, a galaxy which
had collected all of its black holes in the center would show an inner
Keplerian rotation curve, rather than the observed solid body
rotation. Thus, this case can be excluded by inspection.  Even for a
black hole fraction as low as 0.05, after only two half-orbit
time-scales, at $r_0 /4$ the rotation curve would show an increment of
a factor of 2, which could easily be detected. For larger black hole
fractions, e.g. $0.125$, by only one half-orbit time-scale the rotation
velocity at $r_0 /2$ would show an increase of a factor of 1.4, which
could also be easily detectable. We take the maximum $\gamma$ for which
this approach should be valid as $1/8$, at which fraction after 1 half orbit time
the black hole component would concentrate interior to $r_{o}/2$, with an
average density equal to that of the background halo density. Higher black
hole fractions are not applicable, as this simple approach would predict
dynamical friction would concentrate the black hole component to densities
higher than the background halo densities, which is not physical.

We define the critical mass above which black holes can
be ruled out, at a given fraction, as $M_c (\gamma)$, from equation
(12) to obtain:
$$
   M_c (\gamma) ={V_0 r_0^2 n \over 417 (1-\gamma)}   10^5 M_{\odot}
   \eqno\stepeq 
$$
where n is the number of half-orbit time-scales which have elapsed
since the formation of the system. 

For a typical dwarf galaxy we can take $V_0 = 30 km/s$ and $r_0 = 2.5 kpc$,
if we evaluate equation (13) for $n=1$ and $\gamma=1/8$ we obtain
$M_{c}=0.5 \times 10^5 M_{\odot}$. For black holes of greater mass,
the distortion to the solid body rotation curve after one half life had
passed would become noticeable. This shows that black holes
forming less than $1/8$ the mass of dwarf galactic halos, with
individual masses of more than $0.5 \times 10^5 M_{\odot}$ can be
excluded. Lowering the fraction on the halo which the black holes constitute, or
requiring 2 half lives to have elapsed before the effects are noticeable
only introduces a factor of about two,
making the limit mass $M_{c}=1 \times 10^5 M_{\odot}$.
Taking now $V_0 =15 km/s$, and $r_0 =1.0 kpc$, representative of 
a dSph galaxy, we obtain $M_{c}=4 \times 10^3 M_{\odot}$. Again, the
uncertainties in the other parameters in equation (13) introduce a factor
of 2 in this number. From this we see that black holes forming a fraction of 
less than $1/8$ of the dark halos of dSph galaxies can be ruled out, for
masses greater than $1 \times 10^4 M_{\odot}$. Ruling out such small
fractions is not uninteresting, as it is here where other more direct
methods become insensitive.

\section{Conclusions}

From the analysis introduced in section 2 and the
applications of section 3, we can conclude the following:

1) The inward spiraling of a massive body orbiting within the core
region of a dark halo due to dynamical friction proceeds in general at
a different rate than in the isothermal case. Orbital decay in the
constant density case is rapid at first, and slows down as time
progresses, rather than starting slowly and accelerating with time, as
in the isothermal case.

2) The dynamical friction decay time in a constant density dark halo
core, presented in equation (10), can be applied to several problems
involving dynamical friction within dwarf galaxies.  The simple
analytic nature and generality of this expression allows a clear
understanding of the physics and of the scaling properties of 
dynamical friction in this case.

3) Determination and analysis of the spatial distribution 
of the globular cluster systems around dwarf galaxies
can provide insight not only into the evolutionary history of these
systems, but also into the structure of their dark halos. 
In the specific example of the Sagittarius dwarf spheroidal galaxy,
the existence and spatial distribution of its globular cluster system
provides direct evidence for substantial tidal stripping during the
lifetime of the galaxy, as well as making an isothermal halo profile
seem unlikely.

4) The observed smoothness of inner rotation curves in dwarf galaxies,
together with the analysis here, tightly constrains any minority
contribution to their dark halo from individual, high-mass, objects.
Massive black holes with masses of more than $10^4 M_{\odot}$
forming a fraction of less than about 0.1 of the dark halos of dwarf
galaxies can be excluded, as such a contribution to the mass
distribution would evolve by dynamical friction to generate observable
distortions to the rotation curves.

\section*{Acknowledgments}

The work of X. Hernandez was partly supported by a DGAPA-UNAM grant.

\appendix

\section{Solving for the detailed orbital evolution}

From equation (4) and equation (8),

$$
{dr \over dt}=- 16 \pi^2 {ln \Lambda G^2 M \rho_0 r I(X) \over V_c^3}
$$
substituting $V_c$ for $\rho_0$, $V_0$ for $V_c$ and X and $r_0$ for r, we get:

$$
   {dX \over dt}= -A {I(X) \over X^2}, \eqno\stepeq
$$
where:
$$
A=(2 \times 3^5)^{1/2} {\pi G ln \Lambda M \over r_0^3} ({r_0 \over V_0})
 = 277.3 {M \over r_0^3}({r_0 \over V_0})  Gyr^{-1}
$$
with $[V_0]=km/s$, $[r_0]=kpc$ and $[M]=10^5 M_{\odot}$. 

Solving Eq(A1) numerically, one obtains:
$$
   12 \pi^{3/2}(a_1 ln X+a_2 X^2 +a_3 X^3 +...) =-At +C    \eqno\stepeq
$$
Where C is given by the initial conditions, X at t=0, and
$a_1=0.25281$, $a_2=0.07811$, $a_3=0.01079$ and so on.
Notice that $12 \pi^{3/2}a_1=16.89$, which we already knew from F(X=0)=16.89.

In this way, equation (A2) can be used to trace the exact temporal evolution of
the orbit of any massive body, within the core region of a King sphere. 

\section{The effect of the matter distribution beyond $r_0$} 

To explore the dependence of the solution on having neglected the
matter content beyond the core radius, in this subsection we calculate
the escape velocity, $v_e(r)$ considering an exponential cut-off
starting at the core radius, i.e.,

$$
   \rho (r) =\left\{ \matrix{\rho_0 & r<r_0 \cr \rho_0 e^{-(R-1)} &
   r\ge r_0} \right\},  \eqno\stepeq  
$$

Which implies,
$$
{1 \over 2} v_e^2 (r)={4 \pi \over 3} G \rho_0 \left( \int_{r}^{r_0} r
dr +r_0^3 \int_{r_0}^{\infty} {dr \over r} \right)
$$

$$
+ 4 \pi G \rho_0 r_0^2  \int_{1}^{\infty} \left( {5 \over R^2} -{2 \over R^2}
e^{-(R-1)} -{2 \over R}e^{-(R-1)}-e^{-(R-1)} 
\right) dR 
$$
which gives,
$$
{1 \over 2} v_e^2 (r)={4 \pi \over 3} G \rho_0 \left( {r_0^2 -r^2
\over 2} +r_0^2 +6r_0^2  \right) 
$$

$$
   \Longrightarrow  v_e^2(r) = V_c^2 (r) \left( {15 r_0^2 \over r^2}
   -1 \right),  \eqno\stepeq 
$$
which is the equivalent of equation (6), and reflects a higher escape
velocity. This leads to $Y^2=(45/2-X^2)$, which makes the equivalent
of equation (7) become:

$$
   I'(X)=
$$

$$
\hskip 8mm {1 \over 4 \pi} \left( erf(X) -{2X \over \pi^{1/2}}e^{-X^2}
-{4X^3 \over 3 \pi^{1/2}} e^{-(45/2-X)}    \right) \eqno\stepeq 
$$
and 
$$
   \tau =\left( {V_0 \over r_0} \right) \left( {2 \over 3}
   \right)^{3/2} {r_0^3 \over 24 \pi ln \Lambda M G}  F'(X),  \eqno\stepeq
$$
where $F'(X)=X^3/I'(X)$.

We find, $0 <(F(X)-F'(X))<1.65$, for $0<X<1$, with the difference
between the two expressions decreasing rapidly as $X \rightarrow 0$,
and increasing monotonically as $X$ increases, as illustrated in Fig 1. Therefore,
considering a gradual cut-off in the density distribution, even a
quite abrupt exponential one, increases the escape velocity within the
core region enough to reduce the third term in $I(X)$ effectively to
zero. This has the effect of lowering $I(X)$ towards its value at
$X=0$, driving the solution closer to the exponential case, with the
same half-life. This is what one might have expected, as the matter
distribution exterior to the orbiting body does not
enter into the dynamical friction problem, except in as much as it
determines the global distribution function for the halo particles
i.e. $\sigma$.  This dependence has already been taken into account in
the expression for $r_0$ and the assumption of a global King
distribution, and hence one should see no further dependence of the
solution on the matter distribution exterior to the core region. The
only approximation enters in considering the density within the core
region of the King sphere as constant. In as much as the matter
distribution beyond this region does not affect the dynamical friction
problem, the distribution function, circular velocity and density
profile of the halo model are self consistent.

Solving equation (B4) numerically, one gets an identical expansion to
equation (A2), with coefficients: $a_1=0.25$, $a_2=0.075$,
$a_3=0.0091$, etc. Notice that $12 \pi^{3/2} a_1 =16.70$, which is
consistent with $F'(0)=16.70$ and not far from $F(0)=16.89$.

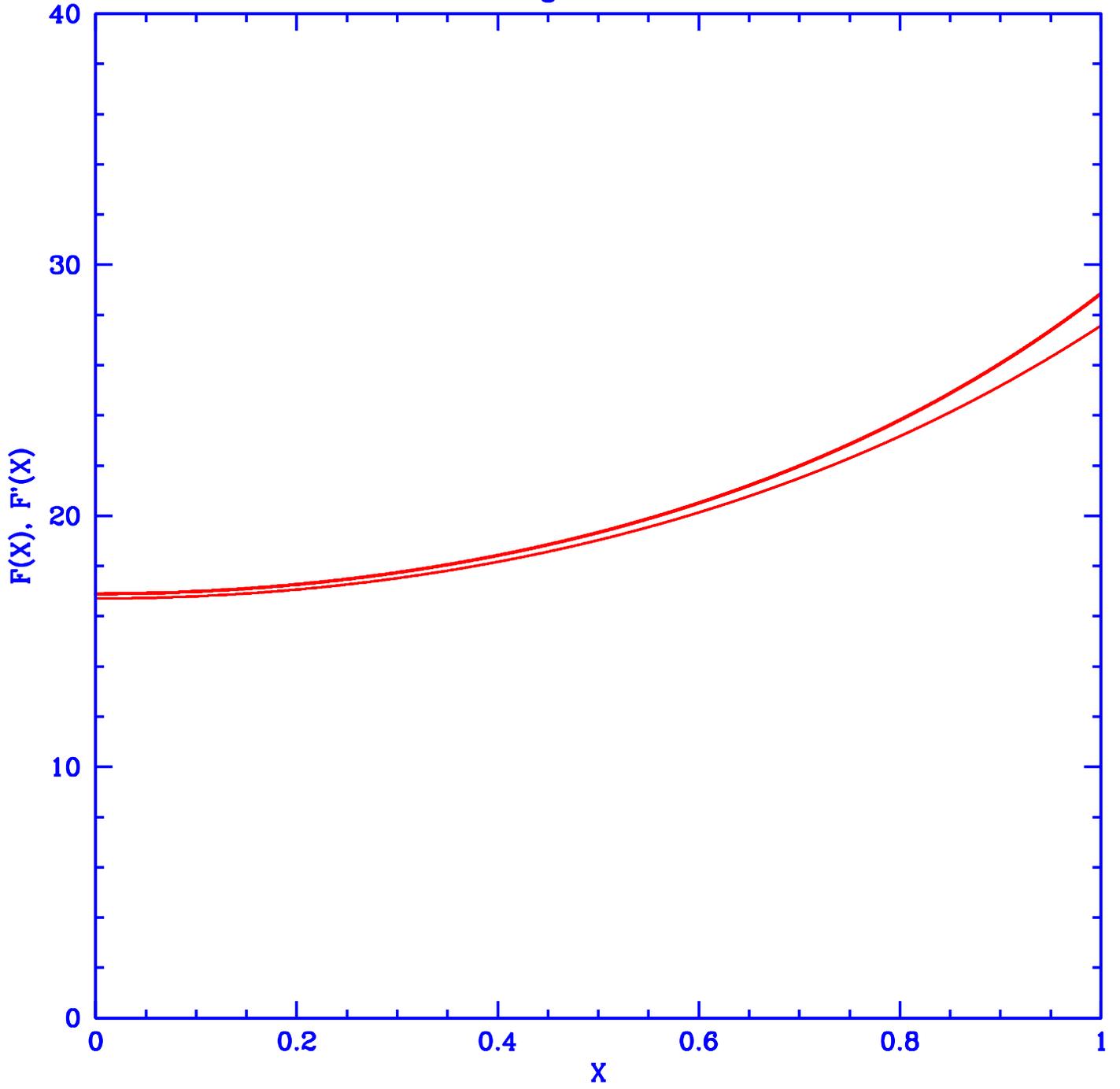
\bye